\documentclass[aps,showpacs,twocolumn,amsmath,amssymb,superscriptaddress,longbibliography]{revtex4-1}

\usepackage{graphicx}
\usepackage{subfigure}
\usepackage{grffile}
\usepackage{bm}
\usepackage{mhchem}
\usepackage{color}
\usepackage{hyperref}

\newcommand{\abs}[1]{\ensuremath{\left| #1 \right|}}
\newcommand{\ket}[1]{{\vert #1\rangle}}

\newcommand{\braket}[2]{\langle#1\vert#2\rangle}
\newcommand{\eval}[3]{\langle#1\vert#2\vert#3\rangle}
\newcommand{\vev}[1]{\langle #1\rangle}
\newcommand{\1}{\mbox{\bf 1}}

\newcommand{\RE}{\text{Re}}

\begin{document}

\title{Photoinduced charge carrier dynamics in Hubbard two-leg ladders and chains}

\author{Can Shao}
\affiliation{Center for Interdisciplinary Studies $\&$ Key Laboratory for Magnetism and Magnetic Materials of the MoE, Lanzhou University, Lanzhou 730000, China}
\affiliation{Beijing Computational Science Research Center, Beijing 100084, China}

\author{Takami Tohyama}
\email{tohyama@rs.tus.ac.jp}
\affiliation{Department of Applied Physics, Tokyo University of Science, Tokyo 125-8585, Japan}

\author{Hong-Gang Luo}
\affiliation{Center for Interdisciplinary Studies $\&$ Key Laboratory for Magnetism and Magnetic Materials of the MoE, Lanzhou University, Lanzhou 730000, China}
\affiliation{Beijing Computational Science Research Center, Beijing 100084, China}

\author{Hantao Lu}
\email{luht@lzu.edu.cn}
\affiliation{Center for Interdisciplinary Studies $\&$ Key Laboratory for Magnetism and Magnetic Materials of the MoE, Lanzhou University, Lanzhou 730000, China}

\date{\today}

\begin{abstract}
The charge carrier dynamics of doped electronic correlated systems on ladders and chains, subject to ultrafast photoirradiation, is investigated using the time-dependent Lanczos method. The time-resolved optical conductivity and the temporal profiles of other relevant quantities, including the doublon number, the kinetic energy, and the interaction energy, are calculated. Two competitive factors that can influence the transient charge carrier dynamics are identified as the thermal effect and the charge effect. We demonstrate that the analysis of their interplay can provide an intuitive way to understand the numerical results and the recent optical pump-probe experiment on a two-leg ladder cuprate.
\end{abstract}


\maketitle


\section{Introduction}\label{sec1}

Time-resolved ultrafast spectroscopy opens a new vista for the study of correlated materials, where an additional dimension on top of energy and momentum is introduced with the intention of disentangling different degrees of freedom in a real-time domain~\cite{koshihara2006special,orenstein2012,aoki2014}. Messages extracted from out-of-equilibrium dynamics are expected to provide irreplaceable information for some key questions such as high-temperature superconductivity in strongly correlated systems~\cite{orenstein2012}. Significant progress has been made on various low-dimensional correlated systems, e.g., on one-dimensional (1D) organic Mott insulators~\cite{iwai2003,okamoto2007,wall2011,mitrano2014} and on cuprates~\cite{ogasawara2000,fausti2011,giannetti2011,conte2012,sentef2013,conte2014,hu2014}. However, due to the highly intertwined quantum fluctuations inherent in correlated states and the nontrivial nature of nonequilibrium processes, it is usually difficult to provide precise and clear theoretical understanding of what has been observed in experiments, and therefore many open questions remain.


Recently, it was reported that on the hole-doped two-leg ladder cuprate Sr$_{14-x}$Ca$_x$Cu$_{24}$O$_{41}$ (SCCO)~\cite{mccarron1988,siegrist1988,uehara1996}, the conductivity is suppressed by applying ultrashort laser pulses within picosecond timescales, while for the undoped mother compound Sr$_{14}$Cu$_{24}$O$_{41}$ (SCO) which is an insulator, a photoinduced metalliclike state has been observed~\cite{fukaya2015}. It was suggested that the change in conductivity induced by the ultrafast photoirradiation in the doped case can be attributed to the disturbance on the coherence of the inherent hole pairs~\cite{fukaya2015}. The argument is supported by numerical studies of the time-dependent pair-field correlation functions on the two-leg Hubbard clusters~\cite{fukaya2015,hashimoto2016}.

Our study is partially stimulated by the above-mentioned interesting observations. The purpose of the present paper is to readdress the charge carrier dynamics of the doped two-leg Hubbard model in the pump-probe process. The impacts of the pumping pulse on the subsequent nonequilibrium charge dynamics can be categorized into two different yet related issues. One is to shake up the existing charges, leading to an increase of the kinetic energy.
We call it the thermal effect. The other factor is to create new charge excitations, including photoinduced itinerant charge carriers. We call it the charge effect. Each factor has a distinctive influence on the Drude weight, the criterion that distinguishes metals from insulators~\cite{kohn1964}.


There have been a number of theoretical works to address the photoinduced insulator-to-metal transitions in the Mott insulators in terms of the Hubbard models in various spatial dimensions~\cite{oka2005,koshihara2006special,takahashi2008,sensarma2010,oka2012,eckstein2013,lu2015,eckstein2016}. Important distinctions between the photo doped and chemically doped cases have been identified. The relaxation of the photoinduced charges has been discussed in detail (e.g., see Refs.~\onlinecite{eckstein2013,sensarma2010,eckstein2016}) for various decay channels of the photoinduced metallic states in the Mott insulators. Nevertheless, few discussions have been devoted to the doped Hubbard(-like) models subjected to ultrafast photoirradiation.

In this paper, working numerically on the two-leg ladders and single chains of the Hubbard model with a sizable on-site Coulomb interaction, we clarify the nature of photoinduced charge dynamics in the doped Mott insulators. Two competitive impacts, i.e., the thermal effect and the charge effect, are identified. Their distinctive influence on the transient charge dynamics can be clearly demonstrated by varying the interleg hopping constant in the ladder model [denoted as $t'$ throughout the paper; see Eq.~(\ref{eq:1})]. In the case of small $t'$ (single chains can be regarded as the limit of $t'=0$), we find the enhancement of the Drude weight owing to the lack of an effective decay channel for photoinduced charges. This is attributed to the charge effect. On the other hand, by increasing $t'$ to the same value as the intraleg hopping $t$, the Drude weight drops rapidly from the initial value due to the thermal effect and the fast decay of photoinduced charges. Actually, as will be shown, a similar phenomenon can be observed on the single chain without and with additional nearest-neighbor interaction.

The outline of the rest of the paper is as follows: In Sec.~\ref{sec2}, after introducing the model Hamiltonian for a ladder that is subject to a time-dependent electric field, a description of the numerical method we employ is presented. Some related technical details are also explained. In Sec.~\ref{sec3} we examine the ultrafast charge carrier dynamics of several related systems, including hole-doped ladders and chains. The conclusion is drawn in Sec.~\ref{sec4}.

\section{Model and method}\label{sec2}

The main Hamiltonian we are working on is the Hubbard model on a two-leg ladder lattice:
\begin{eqnarray}
{\cal H}&=&- t \sum_{\vev{ij}\alpha \sigma} \left(c_{i \alpha \sigma}^\dagger c_{j \alpha \sigma} + {\rm H.c.}\right)
- t' \sum_{i \sigma} \left(c_{i 1 \sigma}^\dagger c_{i 2 \sigma} + {\rm H.c.}\right) \nonumber \\
&&+U \sum_{i \alpha} \left(n_{i \alpha \uparrow}-\frac{1}{2}\right) \left(n_{i \alpha \downarrow}-\frac{1}{2}\right),
\label{eq:1}
\end{eqnarray}
where $c_{i \alpha \sigma}^\dagger$ ($c_{i \alpha \sigma}$) is the creation (annihilation) operator of an electron with spin $\sigma$ at the $i$th rung of the $\alpha$ leg, $\alpha=1,2$; $\vev{ij}$ denotes a pair of nearest-neighbor sites along one leg; the number operator in the Hamiltonian is $n_{i \alpha \sigma}=c_{i \alpha \sigma}^\dagger c_{i \alpha \sigma}$; $t$ and $t'$ are the hopping constants along the longitudinal (leg) and transverse (rung) directions, respectively; their relative strength is defined by a ratio denoted as $r_t := t'/t$; and $U (>0)$ is the on-site repulsive Coulomb interaction. In the following, we set $t$ and $t^{-1}$ as energy and time units with the Planck constant $\hbar=1$. The lattice constant, the speed of light, and the elementary charge are all taken to be unity.

To investigate the photoinduced dynamics under a spatially uniform electric field, the temporal gauge is convenient. In the temporal gauge, the scalar potential is always set to be zero, and the electric field is produced by the temporal dependence of the vector potential as ${\bf E}=-\partial{\bf A}/\partial\tau$ (in the following discussions we reserve $\tau$ exclusively for the time variable). More specifically, for an electric field with polarization parallel to the legs we consider here, a Peierls phase is introduced into the longitudinal hopping terms of the Hamiltonian~(\ref{eq:1}):
\begin{equation}
c^{\dagger}_{i\alpha\sigma}c_{j\alpha\sigma}\rightarrow
e^{iA(\tau)}c^{\dagger}_{i\alpha\sigma}c_{j\alpha\sigma}.
\label{eq:2}
\end{equation}
The vector potential $A(\tau)$ for an electric pulse takes the form
\begin{equation}
A(\tau)=A_{0}\,\exp\left[-\left(\tau-\tau_{0}\right)^2/2\tau_{\text{d}}^2\right]\cos\left[\omega_{p}\left(\tau-\tau_{0}\right)\right],
\label{eq:3}
\end{equation}
with a Gaussian-like envelope around $\tau_0$. The amplitude, the central frequency, and the pulse width are characterized by $A_0$, $\omega_p$, and $\tau_{\text{d}}$, respectively. In the hole-doped case, $\omega_p$ of the pump pulse is set to coincide with the energy of the remnant gap~\cite{hashimoto2016} to activate the charge degree of freedom efficiently, while in the half-filling case, it simply takes the value of the optical gap.

In order to trace the evolution of the system, a straightforward time-dependent Lanczos method is employed as long as the system size is within the reach of the exact diagonalization method. The key formula to evaluate the time-dependent wavefunction $|\psi(\tau)\rangle$ is read
\begin{equation}
\ket{\psi(\tau+\delta{\tau})}\simeq\sum_{l=1}^{M}{e^{-i\epsilon_l\delta{\tau}}}\ket{\phi_l}\braket{\phi_l}{\psi(\tau)},
\label{eq:4}
\end{equation}
where $\epsilon_l$ and $\ket{\phi_l}$ are eigenvalues and eigenvectors of the tridiagonal matrix generated in the Lanczos iteration, respectively; $M$ is the dimension of the Lanczos basis; and $\delta{\tau}$ is the minimum time step. Further details can be found in Ref.~\onlinecite{Prelovsekbook}. In our numerical simulation, we set $M=30$, and $\delta\tau=0.02$. The convergence has been checked by adopting larger $M$ and smaller $\delta\tau$ for comparison.

The essential quantity in our investigation is the time-resolved optical conductivity $\sigma(\omega,\tau)$ (along the longitudinal direction). It is a {\em partial} Fourier transformation of the two-time response function $\sigma(\tau',\tau)$~\cite{kubo1957}, which, in nonequilibrium, is necessary to describe the real-time current response with respect to a time-dependent field $E(t)$. There is however no unique way to perform the partial Fourier transformation~\cite{eckstein2008,eckstein2013,zala2014}. In this paper, we use one of the definitions following Ref.~\onlinecite{zala2014}:
\begin{equation}
\sigma(\omega,\tau)=\int_0^{\infty}\sigma(\tau+s,\tau)e^{i(\omega+i0)s}\,ds.
\label{eq:4b}
\end{equation}

In order to obtain $\sigma(\omega,\tau)$ defined in Eq.~(\ref{eq:4b}), a numerical pump-probe (PP) method is employed. The method has been used to investigate the pump-probe optical spectra of BCS superconductors~\cite{papenkort2007} and the photoinduced in-gap excitations in the 1D Hubbard model~\cite{lu2015}.
More detailed discussions of the method itself can be found in Ref.~\onlinecite{shao2016}. We note that the numerical method we employ here is somehow different from the one used in Refs.~\onlinecite{fukaya2015,hashimoto2016}, where the regular part of the optical conductivity was estimated directly from the current-current correlations with respect to the time-dependent wavefunction $\psi(\tau)$ (e.g., see Eqs. (7) and (8) in Ref.~\onlinecite{hashimoto2016}). We would like to stress that the time-dependent Drude (in-gap) weights obtained by the two methods show consistent behaviors. Our analysis and conclusions do not depend on which method we have employed.

The next issue in the method part is to extract the value of the Drude weight $D(\tau)$ from $\sigma(\omega,\tau)$ with the desired precision. In equilibrium, the Drude weight $D$ appears as the dissipationless component [the coefficient of $\delta(\omega)$] in the real part of the optical conductivity: $\RE\,\sigma(\omega)=2\pi D\delta(\omega)+\RE\,\sigma_{\text{reg}}(\omega)$. Analogously, for the time-resolved optical conductivity in nonequilibrium, the time-dependent Drude weight $D(\tau)$ can be defined as~\cite{zala2014}
\begin{equation}
\RE\,\sigma(\omega,\tau)=2\pi D(\tau)\delta(\omega)+\RE\,\sigma_{\text{reg}}(\omega,\tau).
\label{eq:4c}
\end{equation}
If the kinetic energy contributed by the hoppings along the legs (denoted as $E_t$) and the regular part of the optical conductivity are known, the Drude weight can be readily obtained from the sum rule:
\begin{equation}
D=-\frac{E_t}{2L}-\frac{1}{\pi}\int_0^{\infty}\RE\,\sigma_{\text{reg}}(\omega)\,d\omega,
\label{eq:4d}
\end{equation}
where $D$, $E_t$, and $\RE\,\sigma_{\text{reg}}$ can be temporally dependent, and $L$ is the system size along the current direction. However, we cannot use Eq.~(\ref{eq:4d}) since the numerical PP method we employ is specified to calculate $\sigma(\omega,\tau)$ as a whole~\cite{shao2016}, which contains contributions both from the Drude part and from the regular part.

In order to determine the Drude weight $D(\tau)$ from $\sigma(\omega,\tau)$, we need to find the proper way to figure out the scale of the singularity at $\omega=0$. In the PP method, if the singularity is well separated from the rest of the spectrum, it can be recognized as a peak or dip centered at $\omega=0$ after introducing a broadening factor~\cite{lu2015,shao2016}. It is well known that for the half-filled Hubbard chain with a periodic boundary condition (BC), the Drude weight $D$ exponentially diminishes with the system size. We have reproduced the result by the PP method, which has given us confidence in using the method to evaluate $D$. However, in the present numerical simulation, we prefer the open BC rather than the periodic BC based on the following considerations.

First, note that for the insulating case at half filling the optical conductivity inside the Mott gap vanishes in the open BC, different from the periodic BC where the Drude weight $D$ remains nonzero for a ring (with finite length). On the other hand, at dopings, we have prominent Drude "precursors" (we simply call the Drude precursor the Drude peak and the corresponding finite frequency the Drude frequency) around $\omega\sim 1/L$ in the open BC, which can be recognized as evidence of the existence of the Drude component in the thermodynamic limit~\cite{fye1991}. Second, we observe that the sum rule check on $\RE\,\sigma(\omega,\tau)$ shows a higher degree of accuracy in the open BC compared to the periodic one. It might be due to the translational symmetry breaking in the open BC, which reduces accidental degeneracies in the periodic BC calculation. The conclusion is that working in the open BC has an advantage in the estimation of the Drude component whose value can be more easily extracted and extrapolated to the thermodynamic limit.

We then suggest that in the open BC, the Drude weight can be estimated from the integration of $\RE\,\sigma(\omega,\tau)$ over the low-frequency regime:
\begin{equation}
D(\tau)\propto \int_0^{\omega_c}\RE\,\sigma(\omega,\tau)\,d\omega:=\tilde{D}(\tau),
\label{eq:4e}
\end{equation}
where $\omega_c$ is the cutoff frequency. Regarding the validity of Eq.~(\ref{eq:4e}), there is one key conjecture: We say that in the pump-probe process of the Hubbard chains and ladders, the regular part of $\RE\,\sigma(\omega,\tau)$ is largely confined to the high-energy regime (the same order as the Coulomb interaction $U$, with the condition $U\gg t$ and $t'$). This statement is based upon the observation at equilibrium (zero temperature): At half filling or small dopings with large $U$, $\RE\,\sigma_{\text{reg}}(\omega)$ obtained from the calculation of current-current correlations [see Eq.~(\ref{eq:4f})] becomes perceptible only after some threshold of $\omega_c\sim U$, and there is no observable structure in $\omega\in(0,\omega_c)$~\cite{fye1991}. This can be understood if we recall that
\begin{equation}
\RE\,\sigma_{\text{reg}}(\omega)=\frac{\pi}{L}\sum_{m\neq 0}\frac{\abs{\eval{m}{j}{0}}^2}{E_m-E_0}\delta\left[\omega-\left(E_m-E_0\right)\right],
\label{eq:4f}
\end{equation}
where $j$ is the current operator. At half filling (large $U$), the contribution from finite $\omega$ (i.e., $E_m\neq E_0$) induced by the current $j$ in most cases has to cross some intermediate states with an extra double occupancy. The energy cost is $\omega\sim U$. At small doping there are other minor contributions around $\omega\sim t^2/U$, which only amount to a tiny percentage in the optical spectral weight. These minor contributions are interpreted as the kink excitations due to the motion of the doped charges (holons) in the antiferromagnetic background~\cite{fye1991}. Even in $\RE\,\sigma(\omega,\tau)$, we can identify two well-separated structures: The higher part accommodates the contributions from $\omega\sim U$, and the lower part mainly contains the Drude-type contributions. To be precise, we call the right-hand side of Eq.~(\ref{eq:4f}) with $\omega_c<U$ the in-gap weight, denoted as $\tilde{D}(\tau)$.

\section{Results and discussion}\label{sec3}

Before going into the time-evolved analysis, we would like to first examine the relation between the Drude weight and the pair-field correlations in the ground state of the hole-doped ladder. It has been argued that in the ultrafast photoirradiation on the hole-doped ladder SCCO, there is a strong connection between the transient changes in the Drude component and that of the hole-pair coherence~\cite{fukaya2015}. The later can be estimated by the pair-field functions, which measure the singlet-pairing correlations between rungs. With respect to a given state $\ket{\Psi(\tau)}$, the correlation between the $i$th and $j$th rungs is defined as~\cite{fukaya2015,hashimoto2016}
\begin{equation}
P\left(\abs{i-j},\tau\right)=\eval{\Psi(\tau)}{\Delta_j^\dagger \Delta_i + {\rm H.c.}}{\Psi(\tau)},
\label{eq:5}
\end{equation}
where
\begin{equation}
\Delta_i:=c_{i 2\downarrow}c_{i 1 \uparrow}-c_{i 2 \uparrow}c_{i 1 \downarrow}
\label{eq:6}
\end{equation}
is the pair-field operator on the $i$th rung.

\begin{figure}
\centering
\includegraphics[width=0.46\textwidth,height=0.15\textheight]{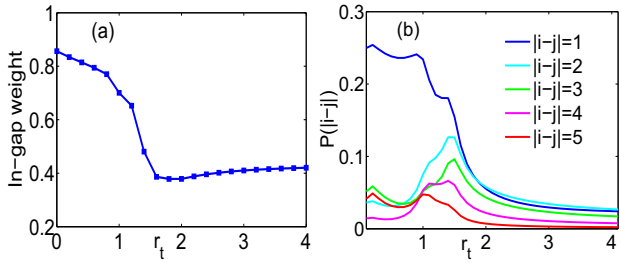}
\caption{(a) The calculated in-gap weight and (b) the pair-field correlation functions vs $r_t$ in the ground state for the ladder Hamiltonian~(\ref{eq:1}) with $1/6$ hole doped in the open BC. Parameters are a $6\times2$ lattice, ten electrons, $U=6$.}
\label{fig1}
\end{figure}

We consider the $1/6$ hole-doped case by using an open $6\times 2$ cluster with five spin-up and five spin-down electrons. The Coulomb interaction $U$ in the Hamiltonian (\ref{eq:1}) is set to be $6$ always. The resulting $r_t$ dependence of the in-gap weight and the pair-field correlations [simply denoted as $P(\abs{i-j})$] at the ground state are shown in Fig.~\ref{fig1}. The in-gap weight is estimated by the integration of $\RE\,\sigma(\omega)$ over the interval $[0,3]$ [Eq.~(\ref{eq:4e})]. The pair-field correlation $P(r)$ as a function of distance is obtained after averaging. That is, for a system size of $6\times 2$, $P(r=1)$ is the average over five available values: $P(i=1,\,j=2)$, $P(i=2,\,j=3)$, $P(i=3,\,j=4)$, $P(i=4,\,j=5)$, and $P(i=5,\,j=6)$; on the other hand, $P(r=5)$ can come from only $P(i=1,\,j=6)$. We note that, as shown in Fig.~\ref{fig1}, the $r_t$ dependence of $P(r>1)$ is quite different from that of the in-gap weight: The prominent pair-field correlations occur at the interval of $r_t\in(1,2)$ but the in-gap weight displays a rapid drop as $r_t$ increases from $0$ towards $2$. As a matter of fact, the nearest-neighbor pairing correlation $P(r=1)$ shows a certain similarity to that of the in-gap weight. The results indicate that the connection between the coherence of hole pairs beyond the nearest neighbor and the Drude weight measured by the in-gap weight is not clearly identified in the ground state, in contrast to the photoexcited states~\cite{fukaya2015}.

\subsection{Hole-doped ladder}

\begin{figure}
\centering
\includegraphics[width=0.46\textwidth,height=0.3\textheight]{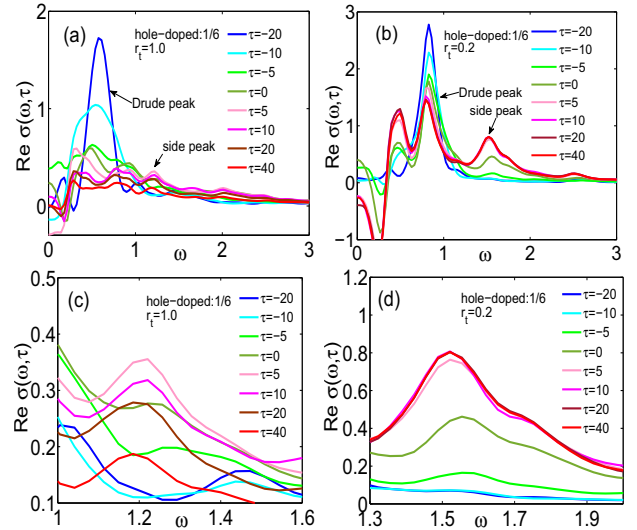}
\caption{The real part of the time-resolved optical conductivity $\RE\,\sigma(\omega,\tau)$ under a pump pulse for the $1/6$ hole-doped open ladder~(\ref{eq:1}) with the anisotropic ratio (a) $r_t=1$ and (b) $r_t=0.2$. $\tau$ is the time delay between the pump and probe pulses. The results for various $\tau$ are shown, including ones before (with negative $\tau$) and after (with positive $\tau$) the pump. (c) and (d) The zoom-in view in the side-peak regime of (a) and (b), respectively. Parameters for the pulse are $A_{0}=0.35$, $\omega_p=4.95$ for $r_t=1$; for $r_t=0.2$, $A_0=0.2$, and $\omega_p=5.64$ (coinciding with the energies of the remnant gap at the given doping). $\tau_{d}=5$ for both cases. Parameters for the model are a $6\times2$ lattice, ten electrons, $U=6$.}
\label{fig2}
\end{figure}

In this section we perform simulations to examine the charge dynamics of the $1/6$ hole-doped ladder under a pump pulse. We first focus on the time-resolved optical conductivity $\RE\,\sigma(\omega,\tau)$. The results for two typical values of $r_t$, i.e., $r_t=1$ and $0.2$, are shown in Fig.~\ref{fig2}. The pumping frequency $\omega_p$ is tuned to $4.95$ and $5.64$, respectively, coinciding with the energy of the remnant gap~\cite{hashimoto2016} to stimulate the charge excitations efficiently. In order to produce equal energy increase for both $r_t$, we use $A_0=0.35$ for $r_t=1$ and $A_0=0.2$ for $r_t=0.2$. Thus, in either case, the energy increment after the pump is around $6.9$.

\begin{figure}
\centering
\includegraphics[width=0.46\textwidth,height=0.3\textheight]{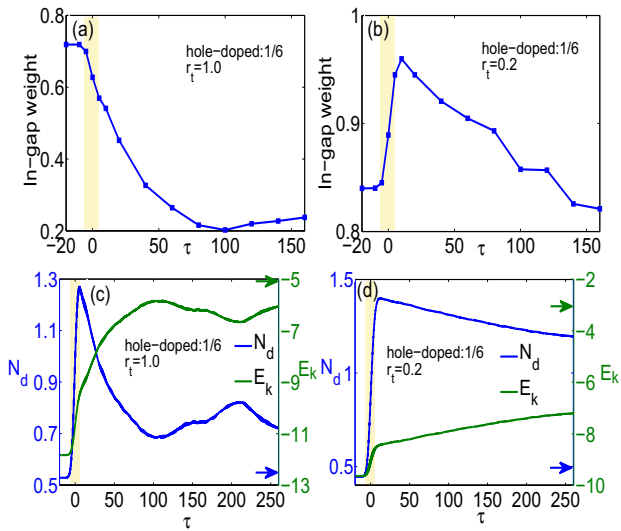}
\caption{(a) (b) The time dependence of the in-gap weight $\tilde{D}$ of the $1/6$ hole-doped open ladder; (c) (d) the temporal profiles (for a longer time interval) of the doublon number $N_d$ and the kinetic energy $E_k$. The settings are identical to Fig.~\ref{fig2}. The shaded areas indicate the short time interval during which the pump pulse is applied. In (c) and (d), the colored arrows on the right-hand sides of the frames indicate the expectation values of $N_d$ and $E_k$ obtained from the microcanonical Lanczos method.}
\label{fig3}
\end{figure}

One of the noticeable features in the time-resolved optical conductivity $\RE\,\sigma(\omega,\tau)$ is the general suppression of the Drude peak (the main peak in Fig.~\ref{fig2}) after the pump, which is largely due to the direct acceleration of the itinerant charges by the pump pulse and the subsequent increase of the kinetic energy [see also the $E_k$ curves in the shaded area in Figs.~\ref{fig3}(c) and \ref{fig3}(d)]. It is consistent with the general thermal-effect picture that was described in the Introduction.

Nevertheless, besides the suppression of the main peak, an additional structure emerges in the neighboring region with energy slightly higher than the Drude frequency (between $\omega\in[1,2]$). We call it the side peak, and zoom-in views are available in Figs.~\ref{fig2}(c) and \ref{fig2}(d). We see that the behaviors of the side peak are quite distinctive between the two different interleg hoppings. In the isotropic case ($r_t=1$) shown in Fig.~\ref{fig2}(c), a side peak with a small amplitude emerges following the application of the pump; as the pulse drops, it begins to diminish after $\tau=5$ and disappears at $\tau\sim 100$ (not shown). In the case of weak interleg hopping $r_t=0.2$ shown in Fig.~\ref{fig2}(d), a prominent side peak appears with a magnitude about three times larger than that in the $r_t=1$ case. The side peak remains without significant decay, up to the maximal time we have performed (around $300$ time units). We note that negative weight near $\omega=0$ in Figs.~\ref{fig2}(a) and \ref{fig2}(b) is caused by the inverse population of photoexcited states, whose magnitude decreases with increasing $\tau$.

We note that the distinctive behavior of the side peak in the weak and moderate interleg hopping regimes has a direct impact on the time dependence of the in-gap weight $\tilde{D}(\tau)$. Recall that $\tilde{D}(\tau)$ is estimated by the integration of $\RE\,\sigma(\omega,\tau)$ over the interval $[0,\omega_c]$ [see Eq.~(\ref{eq:4e})], and as we have mentioned before, the lower and higher energy parts in $\RE\,\sigma(\omega,\tau)$ are clearly separated. Accordingly, we set the cutoff frequency $\omega_c=3$ for $U=6$ to include the side-peak structure~\footnote{An alternative approach of calculating the Drude weight is to employ the sum rule as indicated by Eq.~(\ref{eq:4d}), except replacing the lower limit of the integral in the equation by $\omega_c$. In this case, $\RE\,\sigma_{\text{reg}}(\omega,\tau)$ can be estimated from the time-dependent current-current correlations~\cite{hashimoto2016}. The results from the two different methods have been checked and found to be consistent.}. For $r_t=1$, as shown in Fig.~\ref{fig3}(a), $\tilde{D}(\tau)$ decreases monotonically up to $\tau=100$ and then shows a small recovery from the dip. For $r_t=0.2$ [Fig.~\ref{fig3}(b)], $\tilde{D}(\tau)$ increases initially before a continuous drop takes place around $\tau\approx 10$. The maximal increment is $0.12$. After $\tau\approx 130$, the value of $\tilde{D}(\tau)$ is already lower than the starting value. We note that the contribution to the in-gap weight below the Drude peak in Fig.~\ref{fig3}(b) is small because of the cancellation between positive and negative weights.

To better understand the interplay of the thermal and charge effects on $\tilde{D}(\tau)$, we calculated the temporal profiles of the doublon number $N_d$ and the kinetic energy $E_k$, as shown in Figs.~\ref{fig3}(c) and \ref{fig3}(d). We note that the photoinduced charge excitations in the Hubbard model can be quantified by the increment of the total doublon number $N_d$ with respect to the initial value, where $N_d:=\sum_{i\alpha}n_{i\alpha\uparrow}n_{i\alpha\downarrow}$. The total energy is conserved after the termination of the pump pulse and can be divided into two parts when the Hamiltonian~(\ref{eq:1}) is involved: the kinetic energy $E_k$ from the (intra- and interleg) hopping terms in Eq.~(\ref{eq:1}) and the interaction energy associated with the $U$ term that is proportional to $N_d$ (up to a constant energy shift). In comparison to Figs.~\ref{fig3}(c) and \ref{fig3}(d), we see that the evolution of the side peak in Figs.~\ref{fig2}(c) and \ref{fig2}(d) follows the temporal profiles of $N_d$: If $N_d$ decays quickly after pump, the side peak diminishes as in the case of $r_t=1$; on the other hand, if the decay of $N_d$ is slow as in the case of $r_t=0.2$, the side peak remains for a long time. It corresponds to the charge effect that we addressed in the Introduction. This is the main contribution to the temporary increase of $\tilde{D}(\tau)$ found for $r_t=0.2$.

A close comparison between the top and bottom panels of Fig.~\ref{fig3} shows that the evolution of the in-gap weight $\tilde{D}(\tau)$ follows the temporal profiles of $N_d$ and $E_k$. For $r_t=1$, the fast decay of $N_d$ after the pump associated with the rapid increase of $E_k$ leads to the drop in $\tilde{D}$. Note that $N_d$ has a rebound around $\tau=100$  and has a peak at $\tau=210$. Such a slow-paced oscillation is due to the finite size of the system. Correspondingly, $\tilde{D}$ shows a small recovery from the dip at $\tau=100$. For $r_t=0.2$, the significant growth of $N_d$ and the relatively moderate increase of the kinetic energy during pumping leads to the enhancement of $\tilde{D}$. However, following the process of the subsequent thermalization, i.e., the conversion of the interaction energy into the kinetic one [see Fig.~\ref{fig3}(d)], the in-gap weight $\tilde{D}$ drops from its high point steadily. Note that the scales for the in-gap weight ($y$ axis) in Figs.~\ref{fig3}(a) and \ref{fig3}(b) are different.

So far we have demonstrated that the evolution of the Drude weight is controlled by the competition between the charge effect and the thermal effect. The photoinduced charges that usually have positive contributions to the Drude weight can be recognized as the side peak. On the other hand, the thermal effect responsible for the suppression of the Drude weight takes place with the increment of the kinetic energy. The increase can be either due to the direct pump stimulus or due to the later energy conversion from the interaction part. We conclude that a key ingredient to keep the enhancement of the Drude weight after photoirradiation is to prevent the weight from afterwards decaying or thermalizing, i.e., the thermal effect.

We can easily comprehend the difference of the dissipation rate from interaction energy to the kinetic one between small and large $r_t$ as demonstrated in Figs.~\ref{fig3}(c) and \ref{fig3}(d). Consider the noninteracting model ($U=0$) first, and suppose there are two identical unrelated chains ($t'=0$) at the beginning. After turning on $t'$, the degenerate bands now split into two distinct bands: the bonding and antibonding ones, whose separation in energy is proportional to $t'$. Turning on $U$ next introduces both intra- and interband scattering events. The efficiency of the interband scattering of the doublon-hole recombination should depend on the relative strength between the band separation and the interaction energy, i.e., $t'/U$. In our calculation, we see that the ladder with $r_t=1$ is much more capable of dissipating the $U$ energy compared to $r_t=0.2$ due to the larger interleg hopping constant.

To further access the heating scenario, we present the expectation values of $N_d$ and $E_k$ obtained with the microcanonical Lanczos method (MCLM)~\cite{long2003,Prelovsekbook} as positions of arrows of different colors in Figs.~\ref{fig3}(c) and \ref{fig3}(d). The MCLM values can be regarded as the canonical thermodynamic averages of the system under a given thermal equilibrium energy. They can be used to measure how far away the nonequilibrium system is from the expected thermal state. We can see that for $r_t=1$, $N_d$ and $E_k$ approach their MCLM values much faster than for $r_t=0.2$, which means a more significant thermal effect for $r_t=1$.

\subsection{Hole-doped chain}

\begin{figure}
\centering
\includegraphics[width=0.46\textwidth,height=0.3\textheight]{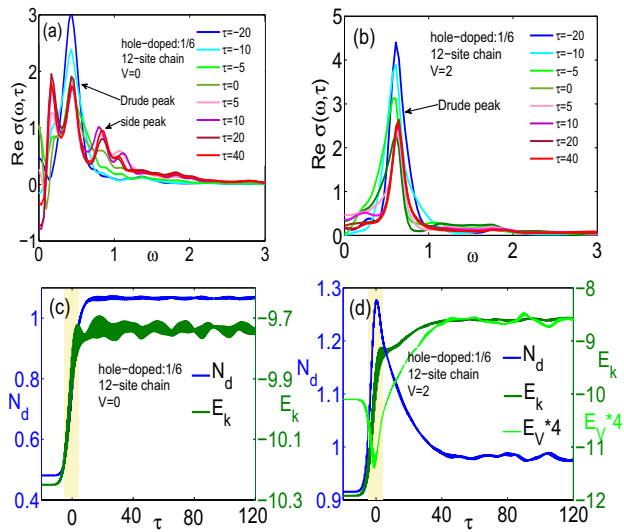}
\caption{The real part of the time-resolved optical conductivity $\RE\,\sigma(\omega,\tau)$ under a pump pulse for the $1/6$ hole-doped open chain~(\ref{eq:7}) without the nearest-neighbor interaction, i.e., (a) $V=0$ and (b) $V=2$. $\tau$ denotes the time delay between the pump and probe pulses. The results for various $\tau$ are shown, including ones before (with negative $\tau$) and after (with positive $\tau$) the pump. The temporal profiles of (c) the doublon number $N_d$ and (d) the kinetic energy $E_k$. Additionally, the evolution of the energy of the $V$ term, denoted as $E_V$, is also plotted in (d) for $V=2$ ($E_V$ is magnified $4$ times for better visualization). Parameters for the pulse are $A_{0}=0.2$ for $V=0$ and $A_0=0.3$ for $V=2$. In both cases, $\tau_d=5$, $\omega_p=5.3$ (the energies of the remnant gap coincide). Parameters for the model are $12$ sites, ten electrons, $U=6$.}
\label{fig4}
\end{figure}

The reasoning mentioned above can be corroborated by further numerical simulations on a single chain, which is regarded as the limit of $r_t=0$. Here we consider the extended Hubbard model on the chain as
\begin{eqnarray}
H&=&-t\sum_{i,\sigma}\left(c_{i,\sigma}^{\dagger}
c_{i+1,\sigma}+{\rm H.c.}\right) \nonumber \\
&&+U\sum_{i}\left(n_{i\uparrow}-\frac{1}{2}\right)
\left(n_{i\downarrow}-\frac{1}{2}\right) \nonumber \\
&&+V\sum_i\left(n_i-1\right)\left(n_{i+1}-1\right),
\label{eq:7}
\end{eqnarray}
with nearest-neighbor interaction $V$. We perform similar calculations on the $12$-site open chain with $1/6$ hole doping (i.e., with ten electrons) as in the case of ladder. We set $A_0=0.2$ for $V=0$ and $A_0=0.3$ for $V=2$ to produce equal energy increases. The energy increments after the pump are both around $4$. The results are summarized in Fig.~\ref{fig4}.

First, let us examine the $U=6, V=0$ case [Figs.~\ref{fig4}(a) and \ref{fig4}(c)]. Around $\tau=0$, accompanied by the reduction of the Drude peak and the appearance of the side peak in $\RE\,\sigma(\omega,\tau)$, as shown in Fig.~\ref{fig4}(a), a surge in both the doublon number $N_d$ and the kinetic energy $E_k$ induced by the pump pulse takes place in Fig.~\ref{fig4}(c). After irradiation, the side peak persists, and $N_d$ and $E_k$ maintain the enhanced value without an observable decrease. Consistent with the behaviors of the side peak, $N_d$ and $E_k$ described above, the in-gap weight is enhanced by the pump pulse and does not show any decrease up to the maximal evolution time we have performed (around $300$ time units; not shown here). This means no thermalization. The situation can be regarded as an extremal case of the ladder with small interleg hoppings. The lack of thermalization is consistent with the expectation from integrable models (e.g., see Ref.~\onlinecite{luca2016} and references therein). On the other hand, for the case of $U=6$ and $V=2$, the side peak is hardly perceived, as shown in Fig.~\ref{fig4}(b). As a consequence, the in-gap weight decreases monotonically. In this case, the total energy is composed of the kinetic energy $E_k$, the $U$ energy quantified by $N_d$, and the nearest-neighbor-interaction energy denoted as $E_V$. Their time evolutions in Fig.~\ref{fig4}(d) indicate that the fast decay of $N_d$ after the pump is mainly due to the restoration of $E_V$. The dissipation into the channel of kinetic energy is quite small.

The study of the chain shows that the doublon-hole recombinations after the pump can take place rapidly with the assistance of the additional $V$ channel, and thus the thermal effect prevails in the decay channel.

Before closing this section, we would like to discuss the analysis of finite size. We performed similar calculations on larger systems of $14$ lattice sites with $12$ electrons (including ladders and chains). Although the doping is slightly different, the results are in good agreement with those of $12$-site clusters (not shown here). Regarding the issue of using the in-gap weight to assess the Drude weight, the arguments are as follows. First, note that the position of the side peak denoted as $\omega_\mathrm{sp}$ is insensitive to $U$. For example, for the ladder with $L=6\times 2$, $r_t=0.2$, and $U=6$, $\omega_\mathrm{sp}=1.52$ [see Fig.~\ref{fig2}(d)]; when $U$ is increased to $10$, $\omega_\mathrm{sp}=1.60$. This indicates that the side peak does not have a magnetic origin. Second, we performed a finite-size scaling of the chains: We calculated $L=6$ with the electron number $N_e=5$, compared to that with $L=12$ and $N_e=10$ (the two size values are the only reasonable ones which can be handled with the Lanczos method at the given doping level). We expect less of a boundary effect for chains compared to ladders. For $U=6$ and $V=0$, we observed that the Drude peak position moves from 0.76 to 0.42 with increasing from $L=6$ to 12. At the same time, the side peak follows the change from $\omega_\mathrm{sp}=1.55$ to 0.82 (with $A_0=0.2$). Note that the distance between the two peaks also decreases with the growth of $L$. Based on these observations, we conclude that the side peak appearing in the optical spectrum after irradiation has similarities to the Drude peak. Although whether the two peaks merge into the singularity at $\omega=0$ in the thermodynamic limit is still an open question in the present study, we consider that the contribution from the side peak should be included in order to estimate the Drude weight correctly.

\subsection{Double pulse on an undoped ladder}

From the previous discussions of the doped systems on ladders and chains, we conclude that a photoenhancement of the Drude weight in the metallic state on ultrafast timescales can be achieved via the generation of specific photoinduced charge excitations, i.e., the charge effect, provided that there are no effective channels in the system for their decay.

\begin{figure}
\centering
\includegraphics[width=0.46\textwidth,height=0.15\textheight]{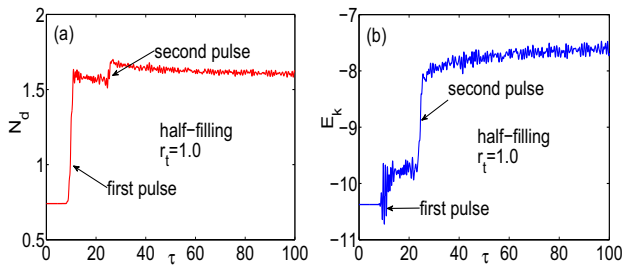}
\caption{(a) The average doublon numbers $N_d$ and (b) the kinetic energy $E_k$ vs time $\tau$ for a double-pulse pumping applied to the ladder Hamiltonian~(\ref{eq:1}) with $r_t=1$ at half filling. The center of the first pulse is located at $\tau=10$, with $A_{0}=0.6$, $\tau_{d}=1$; the second one is located at $\tau=25$, with $A_{0}=0.6$, $\tau_{d}=1$. The central frequencies of the pulses are identical as $\omega_p=3.70$. Parameters for the model are a $6\times2$ lattice, $12$ electrons, $U=6$.}
\label{fig5}
\end{figure}

The competition between the thermal effect and the charge effect can also be demonstrated in a double-pulse setup. We now consider the half-filled ladder Hamiltonian~(\ref{eq:1}) with $r_t=1$ subject to a double photoexcitation. In our simulation, the amplitudes the both pulses are tuned to a large value as $A_0=0.6$. The temporal profiles of the doublon number $N_d$ and the kinetic energy $E_k$ are presented in Fig.~\ref{fig5}.

From Fig.~\ref{fig5}, we observe that the main contribution from the first pulse is to generate charge excitations, which can be vindicated by the surge in $N_d$ around that moment ($\tau=10$). At the same time, the increment of $E_k$ is small since the original state is insulating. This is a typical photoinduced insulator-to-metal transition, and the appearance of the nonzero Drude weight is expected. With the second identical pulse applied after an interval $\Delta\tau=15$, the leading effect is the increment of $E_k$, with little gain with respect to $N_d$. Accordingly the suppression of the Drude weight by the second pump takes place (not shown here). After the double-pulse pumping, a noticeable decay of $N_d$ is clearly observed, accompanied by a further continued increase in $E_k$. We see that the charge effect and the thermal effect dominate in turns in this two-stage evolution.


On the other hand, if we apply relatively weak pump pulses, we observe the enhancement of the Drude weight at both stages~\cite{lu2013}. The reason is that the photoinduced charge excitations with weak irradiation of the first pulse are far from exhausted, leaving plenty of room for further excitations. Thus the charge effect is dominant throughout the double-pulse process. These observations are consistent with the experiment on SCO~\cite{fukaya2015}. Here we note that the fluence dependence of the charge dynamics under the double-pulse pumping presented in Refs.~\onlinecite{fukaya2015,hashimoto2016} can be explained from the point of view of the competition between the charge and thermal effects.

\section{Conclusions}\label{sec4}

To summarize, we carried out numerical simulations on finite-size clusters to investigate the charge carrier dynamics of doped ladders and chains subject to ultrafast optical pumps. We argued that the charge dynamics is led by two competitive factors, namely, the thermal effect and the charge effect. In more detail, the thermal effect includes a shake-up of itinerant charges during irradiation, together with the energy dissipation into the kinetic form in the after-irradiation stage. The effect is characterized by the enhancement of the kinetic energy. On the other hand, the charge effect is associated with the photoinduced charge excitations. A close examination of their respective impacts can lead to an understanding of the charge dynamics of the system under various parameter regimes.

This picture, although quite intuitive, can help us to comprehend the phenomena observed in the numerical simulations and the relevant experimental facts, e.g., of SCCO and its mother compound SCO~\cite{fukaya2015}. We propose that in some materials the photoenhancement of the Drude weight in metallic phases could be observed if the decay channels of induced charge excitations were effectively blocked. For this purpose, it is desired to perform a pump-probe optical measurement for a doped Mott insulator with weakly coupled chains. One possible systems is the cuprate compound PrBa$_2$Cu$_4$O$_8$, in which there is a metallic Cu-O double chain along the $b$ axis, indicating the presence of holons and doublons~\cite{mizokawa2000}. It would be crucial to separate the contribution from the chains. This remains for a future work in pump-probe experiments.


\begin{acknowledgments}
The authors thank J. Bon\v{c}a for helpful discussions.
H.L. and H.-G.L. acknowledge support from the National Natural Science Foundation of China (NSFC; Grants No. 11474136 and No. 11674139) and the Fundamental Research Funds for the Central Universities.
T.T. is partly supported by MEXT, Japan, as a social and scientific priority issue (creation of new functional devices and high-performance materials to support next-generation industries; CDMSI) and exploratory challenge (challenge of basic science -- exploring extremes through multiphysics and multiscale simulations)  on a post-K computer and by CREST (Grant No. JPMJCR1661).
\end{acknowledgments}


\begin{thebibliography}{39}%
\makeatletter
\providecommand \@ifxundefined [1]{%
 \@ifx{#1\undefined}
}%
\providecommand \@ifnum [1]{%
 \ifnum #1\expandafter \@firstoftwo
 \else \expandafter \@secondoftwo
 \fi
}%
\providecommand \@ifx [1]{%
 \ifx #1\expandafter \@firstoftwo
 \else \expandafter \@secondoftwo
 \fi
}%
\providecommand \natexlab [1]{#1}%
\providecommand \enquote  [1]{``#1''}%
\providecommand \bibnamefont  [1]{#1}%
\providecommand \bibfnamefont [1]{#1}%
\providecommand \citenamefont [1]{#1}%
\providecommand \href@noop [0]{\@secondoftwo}%
\providecommand \href [0]{\begingroup \@sanitize@url \@href}%
\providecommand \@href[1]{\@@startlink{#1}\@@href}%
\providecommand \@@href[1]{\endgroup#1\@@endlink}%
\providecommand \@sanitize@url [0]{\catcode `\\12\catcode `\$12\catcode
  `\&12\catcode `\#12\catcode `\^12\catcode `\_12\catcode `\%12\relax}%
\providecommand \@@startlink[1]{}%
\providecommand \@@endlink[0]{}%
\providecommand \url  [0]{\begingroup\@sanitize@url \@url }%
\providecommand \@url [1]{\endgroup\@href {#1}{\urlprefix }}%
\providecommand \urlprefix  [0]{URL }%
\providecommand \Eprint [0]{\href }%
\providecommand \doibase [0]{http://dx.doi.org/}%
\providecommand \selectlanguage [0]{\@gobble}%
\providecommand \bibinfo  [0]{\@secondoftwo}%
\providecommand \bibfield  [0]{\@secondoftwo}%
\providecommand \translation [1]{[#1]}%
\providecommand \BibitemOpen [0]{}%
\providecommand \bibitemStop [0]{}%
\providecommand \bibitemNoStop [0]{.\EOS\space}%
\providecommand \EOS [0]{\spacefactor3000\relax}%
\providecommand \BibitemShut  [1]{\csname bibitem#1\endcsname}%
\let\auto@bib@innerbib\@empty
\bibitem [{\citenamefont {Kuwata-Gonokami}\ and\ \citenamefont
  {(eds)}(2006)}]{koshihara2006special}%
  \BibitemOpen
  {\emph {\bibinfo {booktitle} {Special Topics: Photo-Induced Phase Transitions and their Dynamics}}},\ \bibfield  {author} {\bibinfo {author} {\bibfnamefont {edited by M.}~\bibnamefont
  {Kuwata-Gonokami}}\ and\ \bibinfo {author} {\bibfnamefont {S.~Koshihara}},\ }\href@noop {} {\bibfield  {journal} {\bibinfo
  {journal} {J. Phys. Soc. Jpn.}\ }\textbf {\bibinfo {volume} {75}}{(1)},\ \bibinfo
  {pages} {11001} (\bibinfo {year} {2006})}\BibitemShut {NoStop}%
\bibitem [{\citenamefont {Orenstein}(2012)}]{orenstein2012}%
  \BibitemOpen
  \bibfield  {author} {\bibinfo {author} {\bibfnamefont {J.}~\bibnamefont
  {Orenstein}},\ }\href {\doibase 10.1063/PT.3.1717} {\bibfield  {journal}
  {\bibinfo  {journal} {Phys. Today}\ }\textbf {\bibinfo {volume} {65}},\
  \bibinfo {pages} {44} (\bibinfo {year} {2012})}\BibitemShut {NoStop}%
\bibitem [{\citenamefont {Aoki}\ \emph {et~al.}(2014)\citenamefont {Aoki},
  \citenamefont {Tsuji}, \citenamefont {Eckstein}, \citenamefont {Kollar},
  \citenamefont {Oka},\ and\ \citenamefont {Werner}}]{aoki2014}%
  \BibitemOpen
  \bibfield  {author} {\bibinfo {author} {\bibfnamefont {H.}~\bibnamefont
  {Aoki}}, \bibinfo {author} {\bibfnamefont {N.}~\bibnamefont {Tsuji}},
  \bibinfo {author} {\bibfnamefont {M.}~\bibnamefont {Eckstein}}, \bibinfo
  {author} {\bibfnamefont {M.}~\bibnamefont {Kollar}}, \bibinfo {author}
  {\bibfnamefont {T.}~\bibnamefont {Oka}}, \ and\ \bibinfo {author}
  {\bibfnamefont {P.}~\bibnamefont {Werner}},\ }\href {\doibase
  10.1103/RevModPhys.86.779} {\bibfield  {journal} {\bibinfo  {journal} {Rev.
  Mod. Phys.}\ }\textbf {\bibinfo {volume} {86}},\ \bibinfo {pages} {779}
  (\bibinfo {year} {2014})}\BibitemShut {NoStop}%
\bibitem [{\citenamefont {Iwai}\ \emph {et~al.}(2003)\citenamefont {Iwai},
  \citenamefont {Ono}, \citenamefont {Maeda}, \citenamefont {Matsuzaki},
  \citenamefont {Kishida}, \citenamefont {Okamoto},\ and\ \citenamefont
  {Tokura}}]{iwai2003}%
  \BibitemOpen
  \bibfield  {author} {\bibinfo {author} {\bibfnamefont {S.}~\bibnamefont
  {Iwai}}, \bibinfo {author} {\bibfnamefont {M.}~\bibnamefont {Ono}}, \bibinfo
  {author} {\bibfnamefont {A.}~\bibnamefont {Maeda}}, \bibinfo {author}
  {\bibfnamefont {H.}~\bibnamefont {Matsuzaki}}, \bibinfo {author}
  {\bibfnamefont {H.}~\bibnamefont {Kishida}}, \bibinfo {author} {\bibfnamefont
  {H.}~\bibnamefont {Okamoto}}, \ and\ \bibinfo {author} {\bibfnamefont
  {Y.}~\bibnamefont {Tokura}},\ }\href {\doibase 10.1103/PhysRevLett.91.057401}
  {\bibfield  {journal} {\bibinfo  {journal} {Phys. Rev. Lett.}\ }\textbf
  {\bibinfo {volume} {91}},\ \bibinfo {pages} {057401} (\bibinfo {year}
  {2003})}\BibitemShut {NoStop}%
\bibitem [{\citenamefont {Okamoto}\ \emph {et~al.}(2007)\citenamefont
  {Okamoto}, \citenamefont {Matsuzaki}, \citenamefont {Wakabayashi},
  \citenamefont {Takahashi},\ and\ \citenamefont {Hasegawa}}]{okamoto2007}%
  \BibitemOpen
  \bibfield  {author} {\bibinfo {author} {\bibfnamefont {H.}~\bibnamefont
  {Okamoto}}, \bibinfo {author} {\bibfnamefont {H.}~\bibnamefont {Matsuzaki}},
  \bibinfo {author} {\bibfnamefont {T.}~\bibnamefont {Wakabayashi}}, \bibinfo
  {author} {\bibfnamefont {Y.}~\bibnamefont {Takahashi}}, \ and\ \bibinfo
  {author} {\bibfnamefont {T.}~\bibnamefont {Hasegawa}},\ }\href {\doibase
  10.1103/PhysRevLett.98.037401} {\bibfield  {journal} {\bibinfo  {journal}
  {Phys. Rev. Lett.}\ }\textbf {\bibinfo {volume} {98}},\ \bibinfo {pages}
  {037401} (\bibinfo {year} {2007})}\BibitemShut {NoStop}%
\bibitem [{\citenamefont {Wall}\ \emph {et~al.}(2011)\citenamefont {Wall},
  \citenamefont {Brida}, \citenamefont {Clark}, \citenamefont {Ehrke},
  \citenamefont {Jaksch}, \citenamefont {Ardavan}, \citenamefont {Bonora},
  \citenamefont {Uemura}, \citenamefont {Takahashi}, \citenamefont {Hasegawa},
  \citenamefont {Okamoto}, \citenamefont {Cerullo},\ and\ \citenamefont
  {Cavalleri}}]{wall2011}%
  \BibitemOpen
  \bibfield  {author} {\bibinfo {author} {\bibfnamefont {S.}~\bibnamefont
  {Wall}}, \bibinfo {author} {\bibfnamefont {D.}~\bibnamefont {Brida}},
  \bibinfo {author} {\bibfnamefont {S.~R.}\ \bibnamefont {Clark}}, \bibinfo
  {author} {\bibfnamefont {H.~P.}\ \bibnamefont {Ehrke}}, \bibinfo {author}
  {\bibfnamefont {D.}~\bibnamefont {Jaksch}}, \bibinfo {author} {\bibfnamefont
  {A.}~\bibnamefont {Ardavan}}, \bibinfo {author} {\bibfnamefont
  {S.}~\bibnamefont {Bonora}}, \bibinfo {author} {\bibfnamefont
  {H.}~\bibnamefont {Uemura}}, \bibinfo {author} {\bibfnamefont
  {Y.}~\bibnamefont {Takahashi}}, \bibinfo {author} {\bibfnamefont
  {T.}~\bibnamefont {Hasegawa}}, \bibinfo {author} {\bibfnamefont
  {H.}~\bibnamefont {Okamoto}}, \bibinfo {author} {\bibfnamefont
  {G.}~\bibnamefont {Cerullo}}, \ and\ \bibinfo {author} {\bibfnamefont
  {A.}~\bibnamefont {Cavalleri}},\ }\href {\doibase 10.1038/nphys1831}
  {\bibfield  {journal} {\bibinfo  {journal} {Nat. Phys.}\ }\textbf {\bibinfo
  {volume} {7}},\ \bibinfo {pages} {114} (\bibinfo {year} {2011})}\BibitemShut
  {NoStop}%
\bibitem [{\citenamefont {Mitrano}\ \emph {et~al.}(2014)\citenamefont
  {Mitrano}, \citenamefont {Cotugno}, \citenamefont {Clark}, \citenamefont
  {Singla}, \citenamefont {Kaiser}, \citenamefont {St\"ahler}, \citenamefont
  {Beyer}, \citenamefont {Dressel}, \citenamefont {Baldassarre}, \citenamefont
  {Nicoletti}, \citenamefont {Perucchi}, \citenamefont {Hasegawa},
  \citenamefont {Okamoto}, \citenamefont {Jaksch},\ and\ \citenamefont
  {Cavalleri}}]{mitrano2014}%
  \BibitemOpen
  \bibfield  {author} {\bibinfo {author} {\bibfnamefont {M.}~\bibnamefont
  {Mitrano}}, \bibinfo {author} {\bibfnamefont {G.}~\bibnamefont {Cotugno}},
  \bibinfo {author} {\bibfnamefont {S.~R.}\ \bibnamefont {Clark}}, \bibinfo
  {author} {\bibfnamefont {R.}~\bibnamefont {Singla}}, \bibinfo {author}
  {\bibfnamefont {S.}~\bibnamefont {Kaiser}}, \bibinfo {author} {\bibfnamefont
  {J.}~\bibnamefont {St\"ahler}}, \bibinfo {author} {\bibfnamefont
  {R.}~\bibnamefont {Beyer}}, \bibinfo {author} {\bibfnamefont
  {M.}~\bibnamefont {Dressel}}, \bibinfo {author} {\bibfnamefont
  {L.}~\bibnamefont {Baldassarre}}, \bibinfo {author} {\bibfnamefont
  {D.}~\bibnamefont {Nicoletti}}, \bibinfo {author} {\bibfnamefont
  {A.}~\bibnamefont {Perucchi}}, \bibinfo {author} {\bibfnamefont
  {T.}~\bibnamefont {Hasegawa}}, \bibinfo {author} {\bibfnamefont
  {H.}~\bibnamefont {Okamoto}}, \bibinfo {author} {\bibfnamefont
  {D.}~\bibnamefont {Jaksch}}, \ and\ \bibinfo {author} {\bibfnamefont
  {A.}~\bibnamefont {Cavalleri}},\ }\href {\doibase
  10.1103/PhysRevLett.112.117801} {\bibfield  {journal} {\bibinfo  {journal}
  {Phys. Rev. Lett.}\ }\textbf {\bibinfo {volume} {112}},\ \bibinfo {pages}
  {117801} (\bibinfo {year} {2014})}\BibitemShut {NoStop}%
\bibitem [{\citenamefont {Ogasawara}\ \emph {et~al.}(2000)\citenamefont
  {Ogasawara}, \citenamefont {Ashida}, \citenamefont {Motoyama}, \citenamefont
  {Eisaki}, \citenamefont {Uchida}, \citenamefont {Tokura}, \citenamefont
  {Ghosh}, \citenamefont {Shukla}, \citenamefont {Mazumdar},\ and\
  \citenamefont {Kuwata-Gonokami}}]{ogasawara2000}%
  \BibitemOpen
  \bibfield  {author} {\bibinfo {author} {\bibfnamefont {T.}~\bibnamefont
  {Ogasawara}}, \bibinfo {author} {\bibfnamefont {M.}~\bibnamefont {Ashida}},
  \bibinfo {author} {\bibfnamefont {N.}~\bibnamefont {Motoyama}}, \bibinfo
  {author} {\bibfnamefont {H.}~\bibnamefont {Eisaki}}, \bibinfo {author}
  {\bibfnamefont {S.}~\bibnamefont {Uchida}}, \bibinfo {author} {\bibfnamefont
  {Y.}~\bibnamefont {Tokura}}, \bibinfo {author} {\bibfnamefont
  {H.}~\bibnamefont {Ghosh}}, \bibinfo {author} {\bibfnamefont
  {A.}~\bibnamefont {Shukla}}, \bibinfo {author} {\bibfnamefont
  {S.}~\bibnamefont {Mazumdar}}, \ and\ \bibinfo {author} {\bibfnamefont
  {M.}~\bibnamefont {Kuwata-Gonokami}},\ }\href {\doibase
  10.1103/PhysRevLett.85.2204} {\bibfield  {journal} {\bibinfo  {journal}
  {Phys. Rev. Lett.}\ }\textbf {\bibinfo {volume} {85}},\ \bibinfo {pages}
  {2204} (\bibinfo {year} {2000})}\BibitemShut {NoStop}%
\bibitem [{\citenamefont {Fausti}\ \emph {et~al.}(2011)\citenamefont {Fausti},
  \citenamefont {Tobey}, \citenamefont {Dean}, \citenamefont {Kaiser},
  \citenamefont {Dienst}, \citenamefont {Hoffmann}, \citenamefont {Pyon},
  \citenamefont {Takayama}, \citenamefont {Takagi},\ and\ \citenamefont
  {Cavalleri}}]{fausti2011}%
  \BibitemOpen
  \bibfield  {author} {\bibinfo {author} {\bibfnamefont {D.}~\bibnamefont
  {Fausti}}, \bibinfo {author} {\bibfnamefont {R.~I.}\ \bibnamefont {Tobey}},
  \bibinfo {author} {\bibfnamefont {N.}~\bibnamefont {Dean}}, \bibinfo {author}
  {\bibfnamefont {S.}~\bibnamefont {Kaiser}}, \bibinfo {author} {\bibfnamefont
  {A.}~\bibnamefont {Dienst}}, \bibinfo {author} {\bibfnamefont {M.~C.}\
  \bibnamefont {Hoffmann}}, \bibinfo {author} {\bibfnamefont {S.}~\bibnamefont
  {Pyon}}, \bibinfo {author} {\bibfnamefont {T.}~\bibnamefont {Takayama}},
  \bibinfo {author} {\bibfnamefont {H.}~\bibnamefont {Takagi}}, \ and\ \bibinfo
  {author} {\bibfnamefont {A.}~\bibnamefont {Cavalleri}},\ }\href {\doibase
  10.1126/science.1197294} {\bibfield  {journal} {\bibinfo  {journal}
  {Science}\ }\textbf {\bibinfo {volume} {331}},\ \bibinfo {pages} {189}
  (\bibinfo {year} {2011})}\BibitemShut {NoStop}%
\bibitem [{\citenamefont {Giannetti}\ \emph {et~al.}(2011)\citenamefont
  {Giannetti}, \citenamefont {Cilento}, \citenamefont {Dal~Conte},
  \citenamefont {Coslovich}, \citenamefont {Ferrini}, \citenamefont
  {Molegraaf}, \citenamefont {Raichle}, \citenamefont {Liang}, \citenamefont
  {Eisaki}, \citenamefont {Greven} \emph {et~al.}}]{giannetti2011}%
  \BibitemOpen
  \bibfield  {author} {\bibinfo {author} {\bibfnamefont {C.}~\bibnamefont
  {Giannetti}}, \bibinfo {author} {\bibfnamefont {F.}~\bibnamefont {Cilento}},
  \bibinfo {author} {\bibfnamefont {S.}~\bibnamefont {Dal~Conte}}, \bibinfo
  {author} {\bibfnamefont {G.}~\bibnamefont {Coslovich}}, \bibinfo {author}
  {\bibfnamefont {G.}~\bibnamefont {Ferrini}}, \bibinfo {author} {\bibfnamefont
  {H.}~\bibnamefont {Molegraaf}}, \bibinfo {author} {\bibfnamefont
  {M.}~\bibnamefont {Raichle}}, \bibinfo {author} {\bibfnamefont
  {R.}~\bibnamefont {Liang}}, \bibinfo {author} {\bibfnamefont
  {H.}~\bibnamefont {Eisaki}}, \bibinfo {author} {\bibfnamefont
  {M.}~\bibnamefont {Greven}},  \emph {et~al.},\ }\href {\doibase
  10.1038/ncomms1354} {\bibfield  {journal} {\bibinfo  {journal} {Nat.
  Commun.}\ }\textbf {\bibinfo {volume} {2}},\ \bibinfo {pages} {353} (\bibinfo
  {year} {2011})}\BibitemShut {NoStop}%
\bibitem [{\citenamefont {Dal~Conte}\ \emph {et~al.}(2012)\citenamefont
  {Dal~Conte}, \citenamefont {Giannetti}, \citenamefont {Coslovich},
  \citenamefont {Cilento}, \citenamefont {Bossini}, \citenamefont {Abebaw},
  \citenamefont {Banfi}, \citenamefont {Ferrini}, \citenamefont {Eisaki},
  \citenamefont {Greven}, \citenamefont {Damascelli}, \citenamefont {van~der
  Marel},\ and\ \citenamefont {Parmigiani}}]{conte2012}%
  \BibitemOpen
  \bibfield  {author} {\bibinfo {author} {\bibfnamefont {S.}~\bibnamefont
  {Dal~Conte}}, \bibinfo {author} {\bibfnamefont {C.}~\bibnamefont
  {Giannetti}}, \bibinfo {author} {\bibfnamefont {G.}~\bibnamefont
  {Coslovich}}, \bibinfo {author} {\bibfnamefont {F.}~\bibnamefont {Cilento}},
  \bibinfo {author} {\bibfnamefont {D.}~\bibnamefont {Bossini}}, \bibinfo
  {author} {\bibfnamefont {T.}~\bibnamefont {Abebaw}}, \bibinfo {author}
  {\bibfnamefont {F.}~\bibnamefont {Banfi}}, \bibinfo {author} {\bibfnamefont
  {G.}~\bibnamefont {Ferrini}}, \bibinfo {author} {\bibfnamefont
  {H.}~\bibnamefont {Eisaki}}, \bibinfo {author} {\bibfnamefont
  {M.}~\bibnamefont {Greven}}, \bibinfo {author} {\bibfnamefont
  {A.}~\bibnamefont {Damascelli}}, \bibinfo {author} {\bibfnamefont
  {D.}~\bibnamefont {van~der Marel}}, \ and\ \bibinfo {author} {\bibfnamefont
  {F.}~\bibnamefont {Parmigiani}},\ }\href {\doibase 10.1126/science.1216765}
  {\bibfield  {journal} {\bibinfo  {journal} {Science}\ }\textbf {\bibinfo
  {volume} {335}},\ \bibinfo {pages} {1600} (\bibinfo {year}
  {2012})}\BibitemShut {NoStop}%
\bibitem [{\citenamefont {Sentef}\ \emph {et~al.}(2013)\citenamefont {Sentef},
  \citenamefont {Kemper}, \citenamefont {Moritz}, \citenamefont {Freericks},
  \citenamefont {Shen},\ and\ \citenamefont {Devereaux}}]{sentef2013}%
  \BibitemOpen
  \bibfield  {author} {\bibinfo {author} {\bibfnamefont {M.}~\bibnamefont
  {Sentef}}, \bibinfo {author} {\bibfnamefont {A.~F.}\ \bibnamefont {Kemper}},
  \bibinfo {author} {\bibfnamefont {B.}~\bibnamefont {Moritz}}, \bibinfo
  {author} {\bibfnamefont {J.~K.}\ \bibnamefont {Freericks}}, \bibinfo {author}
  {\bibfnamefont {Z.-X.}\ \bibnamefont {Shen}}, \ and\ \bibinfo {author}
  {\bibfnamefont {T.~P.}\ \bibnamefont {Devereaux}},\ }\href {\doibase
  10.1103/PhysRevX.3.041033} {\bibfield  {journal} {\bibinfo  {journal} {Phys.
  Rev. X}\ }\textbf {\bibinfo {volume} {3}},\ \bibinfo {pages} {041033}
  (\bibinfo {year} {2013})}\BibitemShut {NoStop}%
\bibitem [{\citenamefont {Dal~Conte}\ \emph {et~al.}(2015)\citenamefont
  {Dal~Conte}, \citenamefont {Vidmar}, \citenamefont {Gole{\v{z}}},
  \citenamefont {Mierzejewski}, \citenamefont {Soavi}, \citenamefont {Peli},
  \citenamefont {Banfi}, \citenamefont {Ferrini}, \citenamefont {Comin},
  \citenamefont {Ludbrook}, \citenamefont {Chauviere}, \citenamefont
  {Zhigadlo}, \citenamefont {Eisaki}, \citenamefont {Greven}, \citenamefont
  {Lupi}, \citenamefont {Damascelli}, \citenamefont {Brida}, \citenamefont
  {Capone}, \citenamefont {Bon{\v{c}}a}, \citenamefont {Gerullo},\ and\
  \citenamefont {Giannetti}}]{conte2014}%
  \BibitemOpen
  \bibfield  {author} {\bibinfo {author} {\bibfnamefont {S.}~\bibnamefont
  {Dal~Conte}}, \bibinfo {author} {\bibfnamefont {L.}~\bibnamefont {Vidmar}},
  \bibinfo {author} {\bibfnamefont {D.}~\bibnamefont {Gole{\v{z}}}}, \bibinfo
  {author} {\bibfnamefont {M.}~\bibnamefont {Mierzejewski}}, \bibinfo {author}
  {\bibfnamefont {G.}~\bibnamefont {Soavi}}, \bibinfo {author} {\bibfnamefont
  {S.}~\bibnamefont {Peli}}, \bibinfo {author} {\bibfnamefont {F.}~\bibnamefont
  {Banfi}}, \bibinfo {author} {\bibfnamefont {G.}~\bibnamefont {Ferrini}},
  \bibinfo {author} {\bibfnamefont {R.}~\bibnamefont {Comin}}, \bibinfo
  {author} {\bibfnamefont {B.}~\bibnamefont {Ludbrook}}, \bibinfo {author}
  {\bibfnamefont {L.}~\bibnamefont {Chauviere}}, \bibinfo {author}
  {\bibfnamefont {N.~D.}\ \bibnamefont {Zhigadlo}}, \bibinfo {author}
  {\bibfnamefont {H.}~\bibnamefont {Eisaki}}, \bibinfo {author} {\bibfnamefont
  {M.}~\bibnamefont {Greven}}, \bibinfo {author} {\bibfnamefont
  {S.}~\bibnamefont {Lupi}}, \bibinfo {author} {\bibfnamefont {A.}~\bibnamefont
  {Damascelli}}, \bibinfo {author} {\bibfnamefont {D.}~\bibnamefont {Brida}},
  \bibinfo {author} {\bibfnamefont {M.}~\bibnamefont {Capone}}, \bibinfo
  {author} {\bibfnamefont {J.}~\bibnamefont {Bon{\v{c}}a}}, \bibinfo {author}
  {\bibfnamefont {G.}~\bibnamefont {Gerullo}}, \ and\ \bibinfo {author}
  {\bibfnamefont {C.}~\bibnamefont {Giannetti}},\ }\href {\doibase
  10.1038/nphys3265} {\bibfield  {journal} {\bibinfo  {journal} {Nat. Phys.}\
  }\textbf {\bibinfo {volume} {11}},\ \bibinfo {pages} {421} (\bibinfo {year}
  {2015})}\BibitemShut {NoStop}%
\bibitem [{\citenamefont {Hu}\ \emph {et~al.}(2014)\citenamefont {Hu},
  \citenamefont {Kaiser}, \citenamefont {Nicoletti}, \citenamefont {Hunt},
  \citenamefont {Gierz}, \citenamefont {Hoffmann}, \citenamefont {Le~Tacon},
  \citenamefont {Loew}, \citenamefont {Keimer},\ and\ \citenamefont
  {Cavalleri}}]{hu2014}%
  \BibitemOpen
  \bibfield  {author} {\bibinfo {author} {\bibfnamefont {W.}~\bibnamefont
  {Hu}}, \bibinfo {author} {\bibfnamefont {S.}~\bibnamefont {Kaiser}}, \bibinfo
  {author} {\bibfnamefont {D.}~\bibnamefont {Nicoletti}}, \bibinfo {author}
  {\bibfnamefont {C.~R.}\ \bibnamefont {Hunt}}, \bibinfo {author}
  {\bibfnamefont {I.}~\bibnamefont {Gierz}}, \bibinfo {author} {\bibfnamefont
  {M.~C.}\ \bibnamefont {Hoffmann}}, \bibinfo {author} {\bibfnamefont
  {M.}~\bibnamefont {Le~Tacon}}, \bibinfo {author} {\bibfnamefont
  {T.}~\bibnamefont {Loew}}, \bibinfo {author} {\bibfnamefont {B.}~\bibnamefont
  {Keimer}}, \ and\ \bibinfo {author} {\bibfnamefont {A.}~\bibnamefont
  {Cavalleri}},\ }\href {\doibase 10.1038/nmat3963} {\bibfield  {journal}
  {\bibinfo  {journal} {Nat. Mater.}\ }\textbf {\bibinfo {volume} {13}},\
  \bibinfo {pages} {705} (\bibinfo {year} {2014})}\BibitemShut {NoStop}%
\bibitem [{\citenamefont {McCarron}\ \emph {et~al.}(1988)\citenamefont
  {McCarron}, \citenamefont {Subramanian}, \citenamefont {Calabrese},\ and\
  \citenamefont {Harlow}}]{mccarron1988}%
  \BibitemOpen
  \bibfield  {author} {\bibinfo {author} {\bibfnamefont {E.}~\bibnamefont
  {McCarron}}, \bibinfo {author} {\bibfnamefont {M.}~\bibnamefont
  {Subramanian}}, \bibinfo {author} {\bibfnamefont {J.}~\bibnamefont
  {Calabrese}}, \ and\ \bibinfo {author} {\bibfnamefont {R.}~\bibnamefont
  {Harlow}},\ }\href {\doibase https://doi.org/10.1016/0025-5408(88)90124-9}
  {\bibfield  {journal} {\bibinfo  {journal} {Mater. Res. Bull.}\ }\textbf
  {\bibinfo {volume} {23}},\ \bibinfo {pages} {1355 } (\bibinfo {year}
  {1988})}\BibitemShut {NoStop}%
\bibitem [{\citenamefont {Siegrist}\ \emph {et~al.}(1988)\citenamefont
  {Siegrist}, \citenamefont {Schneemeyer}, \citenamefont {Sunshine},
  \citenamefont {Waszczak},\ and\ \citenamefont {Roth}}]{siegrist1988}%
  \BibitemOpen
  \bibfield  {author} {\bibinfo {author} {\bibfnamefont {T.}~\bibnamefont
  {Siegrist}}, \bibinfo {author} {\bibfnamefont {L.}~\bibnamefont
  {Schneemeyer}}, \bibinfo {author} {\bibfnamefont {S.}~\bibnamefont
  {Sunshine}}, \bibinfo {author} {\bibfnamefont {J.}~\bibnamefont {Waszczak}},
  \ and\ \bibinfo {author} {\bibfnamefont {R.}~\bibnamefont {Roth}},\ }\href
  {\doibase https://doi.org/10.1016/0025-5408(88)90268-1} {\bibfield  {journal}
  {\bibinfo  {journal} {Mater. Res. Bull.}\ }\textbf {\bibinfo {volume} {23}},\
  \bibinfo {pages} {1429 } (\bibinfo {year} {1988})}\BibitemShut {NoStop}%
\bibitem [{\citenamefont {Uehara}\ \emph {et~al.}(1996)\citenamefont {Uehara},
  \citenamefont {Nagata}, \citenamefont {Akimitsu}, \citenamefont {Takahashi},
  \citenamefont {M\^ori},\ and\ \citenamefont {Kinoshita}}]{uehara1996}%
  \BibitemOpen
  \bibfield  {author} {\bibinfo {author} {\bibfnamefont {M.}~\bibnamefont
  {Uehara}}, \bibinfo {author} {\bibfnamefont {T.}~\bibnamefont {Nagata}},
  \bibinfo {author} {\bibfnamefont {J.}~\bibnamefont {Akimitsu}}, \bibinfo
  {author} {\bibfnamefont {H.}~\bibnamefont {Takahashi}}, \bibinfo {author}
  {\bibfnamefont {N.}~\bibnamefont {M\^ori}}, \ and\ \bibinfo {author}
  {\bibfnamefont {K.}~\bibnamefont {Kinoshita}},\ }\href {\doibase
  10.1143/JPSJ.65.2764} {\bibfield  {journal} {\bibinfo  {journal} {J. Phys.
  Soc. Jpn.}\ }\textbf {\bibinfo {volume} {65}},\ \bibinfo {pages} {2764}
  (\bibinfo {year} {1996})}\BibitemShut {NoStop}%
\bibitem [{\citenamefont {Fukaya}\ \emph {et~al.}(2015)\citenamefont {Fukaya},
  \citenamefont {Okimoto}, \citenamefont {Kunitomo}, \citenamefont {Onda},
  \citenamefont {Ishikawa}, \citenamefont {Koshihara}, \citenamefont
  {Hashimoto}, \citenamefont {Ishihara}, \citenamefont {Isayama}, \citenamefont
  {Yui},\ and\ \citenamefont {Sasagawa}}]{fukaya2015}%
  \BibitemOpen
  \bibfield  {author} {\bibinfo {author} {\bibfnamefont {R.}~\bibnamefont
  {Fukaya}}, \bibinfo {author} {\bibfnamefont {Y.}~\bibnamefont {Okimoto}},
  \bibinfo {author} {\bibfnamefont {M.}~\bibnamefont {Kunitomo}}, \bibinfo
  {author} {\bibfnamefont {K.}~\bibnamefont {Onda}}, \bibinfo {author}
  {\bibfnamefont {T.}~\bibnamefont {Ishikawa}}, \bibinfo {author}
  {\bibfnamefont {S.}~\bibnamefont {Koshihara}}, \bibinfo {author}
  {\bibfnamefont {H.}~\bibnamefont {Hashimoto}}, \bibinfo {author}
  {\bibfnamefont {S.}~\bibnamefont {Ishihara}}, \bibinfo {author}
  {\bibfnamefont {A.}~\bibnamefont {Isayama}}, \bibinfo {author} {\bibfnamefont
  {H.}~\bibnamefont {Yui}}, \ and\ \bibinfo {author} {\bibfnamefont
  {T.}~\bibnamefont {Sasagawa}},\ }\href {\doibase 10.1038/ncomms9519}
  {\bibfield  {journal} {\bibinfo  {journal} {Nat. Commun.}\ }\textbf {\bibinfo
  {volume} {6}},\ \bibinfo {pages} {8519} (\bibinfo {year} {2015})}\BibitemShut
  {NoStop}%
\bibitem [{\citenamefont {Hashimoto}\ and\ \citenamefont
  {Ishihara}(2016)}]{hashimoto2016}%
  \BibitemOpen
  \bibfield  {author} {\bibinfo {author} {\bibfnamefont {H.}~\bibnamefont
  {Hashimoto}}\ and\ \bibinfo {author} {\bibfnamefont {S.}~\bibnamefont
  {Ishihara}},\ }\href {\doibase 10.1103/PhysRevB.93.165133} {\bibfield
  {journal} {\bibinfo  {journal} {Phys. Rev. B}\ }\textbf {\bibinfo {volume}
  {93}},\ \bibinfo {pages} {165133} (\bibinfo {year} {2016})}\BibitemShut
  {NoStop}%
\bibitem [{\citenamefont {Kohn}(1964)}]{kohn1964}%
  \BibitemOpen
  \bibfield  {author} {\bibinfo {author} {\bibfnamefont {W.}~\bibnamefont
  {Kohn}},\ }\href {\doibase 10.1103/PhysRev.133.A171} {\bibfield  {journal}
  {\bibinfo  {journal} {Phys. Rev.}\ }\textbf {\bibinfo {volume} {133}},\
  \bibinfo {pages} {A171} (\bibinfo {year} {1964})}\BibitemShut {NoStop}%
\bibitem [{\citenamefont {Oka}\ and\ \citenamefont {Aoki}(2005)}]{oka2005}%
  \BibitemOpen
  \bibfield  {author} {\bibinfo {author} {\bibfnamefont {T.}~\bibnamefont
  {Oka}}\ and\ \bibinfo {author} {\bibfnamefont {H.}~\bibnamefont {Aoki}},\
  }\href {\doibase 10.1103/PhysRevLett.95.137601} {\bibfield  {journal}
  {\bibinfo  {journal} {Phys. Rev. Lett.}\ }\textbf {\bibinfo {volume} {95}},\
  \bibinfo {pages} {137601} (\bibinfo {year} {2005})}\BibitemShut {NoStop}%
\bibitem [{\citenamefont {Takahashi}\ \emph {et~al.}(2008)\citenamefont
  {Takahashi}, \citenamefont {Itoh},\ and\ \citenamefont
  {Aihara}}]{takahashi2008}%
  \BibitemOpen
  \bibfield  {author} {\bibinfo {author} {\bibfnamefont {A.}~\bibnamefont
  {Takahashi}}, \bibinfo {author} {\bibfnamefont {H.}~\bibnamefont {Itoh}}, \
  and\ \bibinfo {author} {\bibfnamefont {M.}~\bibnamefont {Aihara}},\ }\href
  {\doibase 10.1103/PhysRevB.77.205105} {\bibfield  {journal} {\bibinfo
  {journal} {Phys. Rev. B}\ }\textbf {\bibinfo {volume} {77}},\ \bibinfo
  {pages} {205105} (\bibinfo {year} {2008})}\BibitemShut {NoStop}%
\bibitem [{\citenamefont {Sensarma}\ \emph {et~al.}(2010)\citenamefont
  {Sensarma}, \citenamefont {Pekker}, \citenamefont {Altman}, \citenamefont
  {Demler}, \citenamefont {Strohmaier}, \citenamefont {Greif}, \citenamefont
  {J\"ordens}, \citenamefont {Tarruell}, \citenamefont {Moritz},\ and\
  \citenamefont {Esslinger}}]{sensarma2010}%
  \BibitemOpen
  \bibfield  {author} {\bibinfo {author} {\bibfnamefont {R.}~\bibnamefont
  {Sensarma}}, \bibinfo {author} {\bibfnamefont {D.}~\bibnamefont {Pekker}},
  \bibinfo {author} {\bibfnamefont {E.}~\bibnamefont {Altman}}, \bibinfo
  {author} {\bibfnamefont {E.}~\bibnamefont {Demler}}, \bibinfo {author}
  {\bibfnamefont {N.}~\bibnamefont {Strohmaier}}, \bibinfo {author}
  {\bibfnamefont {D.}~\bibnamefont {Greif}}, \bibinfo {author} {\bibfnamefont
  {R.}~\bibnamefont {J\"ordens}}, \bibinfo {author} {\bibfnamefont
  {L.}~\bibnamefont {Tarruell}}, \bibinfo {author} {\bibfnamefont
  {H.}~\bibnamefont {Moritz}}, \ and\ \bibinfo {author} {\bibfnamefont
  {T.}~\bibnamefont {Esslinger}},\ }\href {\doibase 10.1103/PhysRevB.82.224302}
  {\bibfield  {journal} {\bibinfo  {journal} {Phys. Rev. B}\ }\textbf {\bibinfo
  {volume} {82}},\ \bibinfo {pages} {224302} (\bibinfo {year}
  {2010})}\BibitemShut {NoStop}%
\bibitem [{\citenamefont {Oka}(2012)}]{oka2012}%
  \BibitemOpen
  \bibfield  {author} {\bibinfo {author} {\bibfnamefont {T.}~\bibnamefont
  {Oka}},\ }\href {http://prb.aps.org/abstract/PRB/v86/i7/e075148} {\bibfield
  {journal} {\bibinfo  {journal} {Phys. Rev. B}\ }\textbf {\bibinfo {volume}
  {86}},\ \bibinfo {pages} {075148} (\bibinfo {year} {2012})}\BibitemShut
  {NoStop}%
\bibitem [{\citenamefont {Eckstein}\ and\ \citenamefont
  {Werner}(2013)}]{eckstein2013}%
  \BibitemOpen
  \bibfield  {author} {\bibinfo {author} {\bibfnamefont {M.}~\bibnamefont
  {Eckstein}}\ and\ \bibinfo {author} {\bibfnamefont {P.}~\bibnamefont
  {Werner}},\ }\href {\doibase 10.1103/PhysRevLett.110.126401} {\bibfield
  {journal} {\bibinfo  {journal} {Phys. Rev. Lett.}\ }\textbf {\bibinfo
  {volume} {110}},\ \bibinfo {pages} {126401} (\bibinfo {year}
  {2013})}\BibitemShut {NoStop}%
\bibitem [{\citenamefont {Lu}\ \emph {et~al.}(2015)\citenamefont {Lu},
  \citenamefont {Shao}, \citenamefont {Bon\ifmmode~\check{c}\else \v{c}\fi{}a},
  \citenamefont {Manske},\ and\ \citenamefont {Tohyama}}]{lu2015}%
  \BibitemOpen
  \bibfield  {author} {\bibinfo {author} {\bibfnamefont {H.}~\bibnamefont
  {Lu}}, \bibinfo {author} {\bibfnamefont {C.}~\bibnamefont {Shao}}, \bibinfo
  {author} {\bibfnamefont {J.}~\bibnamefont {Bon\ifmmode~\check{c}\else
  \v{c}\fi{}a}}, \bibinfo {author} {\bibfnamefont {D.}~\bibnamefont {Manske}},
  \ and\ \bibinfo {author} {\bibfnamefont {T.}~\bibnamefont {Tohyama}},\ }\href
  {\doibase 10.1103/PhysRevB.91.245117} {\bibfield  {journal} {\bibinfo
  {journal} {Phys. Rev. B}\ }\textbf {\bibinfo {volume} {91}},\ \bibinfo
  {pages} {245117} (\bibinfo {year} {2015})}\BibitemShut {NoStop}%
\bibitem [{\citenamefont {Eckstein}\ and\ \citenamefont
  {Werner}(2016)}]{eckstein2016}%
  \BibitemOpen
  \bibfield  {author} {\bibinfo {author} {\bibfnamefont {M.}~\bibnamefont
  {Eckstein}}\ and\ \bibinfo {author} {\bibfnamefont {P.}~\bibnamefont
  {Werner}},\ }\href {http://dx.doi.org/10.1038/srep21235
  http://www.nature.com/articles/srep21235} {\bibfield  {journal} {\bibinfo
  {journal} {Sci. Rep.}\ }\textbf {\bibinfo {volume} {6}},\ \bibinfo {pages}
  {21235} (\bibinfo {year} {2016})}\BibitemShut {NoStop}%
\bibitem [{\citenamefont {Prelov{\v{s}}ek}\ and\ \citenamefont
  {Bon{\v{c}}a}(2013)}]{Prelovsekbook}%
  \BibitemOpen
  \bibfield  {author} {\bibinfo {author} {\bibfnamefont {P.}~\bibnamefont
  {Prelov{\v{s}}ek}}\ and\ \bibinfo {author} {\bibfnamefont {J.}~\bibnamefont
  {Bon{\v{c}}a}},\ }in\ \href@noop {} {\emph {\bibinfo {booktitle} {Strongly
  Correlated Systems}}},\ \bibinfo {series} {Springer Series in Solid-State
  Sciences}, Vol.\ \bibinfo {volume} {176},\ \bibinfo {editor} {edited by\
  \bibinfo {editor} {\bibfnamefont {A.}~\bibnamefont {Avella}}\ and\ \bibinfo
  {editor} {\bibfnamefont {F.}~\bibnamefont {Mancini}}}\ (\bibinfo  {publisher}
  {Springer, Berlin},\ \bibinfo {year} {2013})\ pp.\ \bibinfo {pages}
  {1--30}\BibitemShut {NoStop}%
\bibitem [{\citenamefont {Kubo}(1957)}]{kubo1957}%
  \BibitemOpen
  \bibfield  {author} {\bibinfo {author} {\bibfnamefont {R.}~\bibnamefont
  {Kubo}},\ }\href@noop {} {\bibfield  {journal} {\bibinfo  {journal} {J. Phys.
  Soc. Jpn.}\ }\textbf {\bibinfo {volume} {12}},\ \bibinfo {pages} {570}
  (\bibinfo {year} {1957})}\BibitemShut {NoStop}%
\bibitem [{\citenamefont {Eckstein}\ and\ \citenamefont
  {Kollar}(2008)}]{eckstein2008}%
  \BibitemOpen
  \bibfield  {author} {\bibinfo {author} {\bibfnamefont {M.}~\bibnamefont
  {Eckstein}}\ and\ \bibinfo {author} {\bibfnamefont {M.}~\bibnamefont
  {Kollar}},\ }\href {\doibase 10.1103/PhysRevB.78.205119} {\bibfield
  {journal} {\bibinfo  {journal} {Phys. Rev. B}\ }\textbf {\bibinfo {volume}
  {78}},\ \bibinfo {pages} {205119} (\bibinfo {year} {2008})}\BibitemShut
  {NoStop}%
\bibitem [{\citenamefont {Lenar\ifmmode \check{c}\else
  \v{c}\fi{}i\ifmmode~\check{c}\else \v{c}\fi{}}\ \emph
  {et~al.}(2014)\citenamefont {Lenar\ifmmode \check{c}\else
  \v{c}\fi{}i\ifmmode~\check{c}\else \v{c}\fi{}}, \citenamefont
  {Gole\ifmmode~\check{z}\else \v{z}\fi{}}, \citenamefont
  {Bon\ifmmode~\check{c}\else \v{c}\fi{}a},\ and\ \citenamefont
  {Prelov\ifmmode~\check{s}\else \v{s}\fi{}ek}}]{zala2014}%
  \BibitemOpen
  \bibfield  {author} {\bibinfo {author} {\bibfnamefont {Z.}~\bibnamefont
  {Lenar\ifmmode \check{c}\else \v{c}\fi{}i\ifmmode~\check{c}\else
  \v{c}\fi{}}}, \bibinfo {author} {\bibfnamefont {D.}~\bibnamefont
  {Gole\ifmmode~\check{z}\else \v{z}\fi{}}}, \bibinfo {author} {\bibfnamefont
  {J.}~\bibnamefont {Bon\ifmmode~\check{c}\else \v{c}\fi{}a}}, \ and\ \bibinfo
  {author} {\bibfnamefont {P.}~\bibnamefont {Prelov\ifmmode~\check{s}\else
  \v{s}\fi{}ek}},\ }\href {\doibase 10.1103/PhysRevB.89.125123} {\bibfield
  {journal} {\bibinfo  {journal} {Phys. Rev. B}\ }\textbf {\bibinfo {volume}
  {89}},\ \bibinfo {pages} {125123} (\bibinfo {year} {2014})}\BibitemShut
  {NoStop}%
\bibitem [{\citenamefont {Papenkort}\ \emph {et~al.}(2007)\citenamefont
  {Papenkort}, \citenamefont {Axt},\ and\ \citenamefont
  {Kuhn}}]{papenkort2007}%
  \BibitemOpen
  \bibfield  {author} {\bibinfo {author} {\bibfnamefont {T.}~\bibnamefont
  {Papenkort}}, \bibinfo {author} {\bibfnamefont {V.}~\bibnamefont {Axt}}, \
  and\ \bibinfo {author} {\bibfnamefont {T.}~\bibnamefont {Kuhn}},\ }\href
  {\doibase 10.1103/PhysRevB.76.224522} {\bibfield  {journal} {\bibinfo
  {journal} {Phys. Rev. B}\ }\textbf {\bibinfo {volume} {76}},\ \bibinfo
  {pages} {224522} (\bibinfo {year} {2007})}\BibitemShut {NoStop}%
\bibitem [{\citenamefont {Shao}\ \emph {et~al.}(2016)\citenamefont {Shao},
  \citenamefont {Tohyama}, \citenamefont {Luo},\ and\ \citenamefont
  {Lu}}]{shao2016}%
  \BibitemOpen
  \bibfield  {author} {\bibinfo {author} {\bibfnamefont {C.}~\bibnamefont
  {Shao}}, \bibinfo {author} {\bibfnamefont {T.}~\bibnamefont {Tohyama}},
  \bibinfo {author} {\bibfnamefont {H.-G.}\ \bibnamefont {Luo}}, \ and\
  \bibinfo {author} {\bibfnamefont {H.}~\bibnamefont {Lu}},\ }\href {\doibase
  10.1103/PhysRevB.93.195144} {\bibfield  {journal} {\bibinfo  {journal} {Phys.
  Rev. B}\ }\textbf {\bibinfo {volume} {93}},\ \bibinfo {pages} {195144}
  (\bibinfo {year} {2016})}\BibitemShut {NoStop}%
\bibitem [{\citenamefont {Fye}\ \emph {et~al.}(1991)\citenamefont {Fye},
  \citenamefont {Martins}, \citenamefont {Scalapino}, \citenamefont {Wagner},\
  and\ \citenamefont {Hanke}}]{fye1991}%
  \BibitemOpen
  \bibfield  {author} {\bibinfo {author} {\bibfnamefont {R.~M.}\ \bibnamefont
  {Fye}}, \bibinfo {author} {\bibfnamefont {M.~J.}\ \bibnamefont {Martins}},
  \bibinfo {author} {\bibfnamefont {D.~J.}\ \bibnamefont {Scalapino}}, \bibinfo
  {author} {\bibfnamefont {J.}~\bibnamefont {Wagner}}, \ and\ \bibinfo {author}
  {\bibfnamefont {W.}~\bibnamefont {Hanke}},\ }\href {\doibase
  10.1103/PhysRevB.44.6909} {\bibfield  {journal} {\bibinfo  {journal} {Phys.
  Rev. B}\ }\textbf {\bibinfo {volume} {44}},\ \bibinfo {pages} {6909}
  (\bibinfo {year} {1991})}\BibitemShut {NoStop}%
\bibitem [{Note1()}]{Note1}%
  \BibitemOpen
  \bibinfo {note} {An alternative approach of calculating the Drude weight is
  to employ the sum rule as indicated by Eq.~(\ref {eq:4d}), except replacing
  the lower limit of the integral in the equation by $\omega _c$. In this case,
  $\protect \text {Re}\protect \tmspace +\thinmuskip {.1667em}\sigma _{\protect
  \text {reg}}(\omega ,\tau )$ can be estimated from the time-dependent
  current-current correlations~\cite {hashimoto2016}. The results from the two
  different methods have been checked and found to be consistent.}\BibitemShut
  {Stop}%
\bibitem [{\citenamefont {Long}\ \emph {et~al.}(2003)\citenamefont {Long},
  \citenamefont {Prelov\ifmmode~\check{s}\else \v{s}\fi{}ek}, \citenamefont
  {El~Shawish}, \citenamefont {Karadamoglou},\ and\ \citenamefont
  {Zotos}}]{long2003}%
  \BibitemOpen
  \bibfield  {author} {\bibinfo {author} {\bibfnamefont {M.~W.}\ \bibnamefont
  {Long}}, \bibinfo {author} {\bibfnamefont {P.}~\bibnamefont
  {Prelov\ifmmode~\check{s}\else \v{s}\fi{}ek}}, \bibinfo {author}
  {\bibfnamefont {S.}~\bibnamefont {El~Shawish}}, \bibinfo {author}
  {\bibfnamefont {J.}~\bibnamefont {Karadamoglou}}, \ and\ \bibinfo {author}
  {\bibfnamefont {X.}~\bibnamefont {Zotos}},\ }\href {\doibase
  10.1103/PhysRevB.68.235106} {\bibfield  {journal} {\bibinfo  {journal} {Phys.
  Rev. B}\ }\textbf {\bibinfo {volume} {68}},\ \bibinfo {pages} {235106}
  (\bibinfo {year} {2003})}\BibitemShut {NoStop}%
\bibitem [{\citenamefont {D'Alessio}\ \emph {et~al.}(2016)\citenamefont
  {D'Alessio}, \citenamefont {Kafri}, \citenamefont {Polkovnikov},\ and\
  \citenamefont {Rigol}}]{luca2016}%
  \BibitemOpen
  \bibfield  {author} {\bibinfo {author} {\bibfnamefont {L.}~\bibnamefont
  {D'Alessio}}, \bibinfo {author} {\bibfnamefont {Y.}~\bibnamefont {Kafri}},
  \bibinfo {author} {\bibfnamefont {A.}~\bibnamefont {Polkovnikov}}, \ and\
  \bibinfo {author} {\bibfnamefont {M.}~\bibnamefont {Rigol}},\ }\href
  {\doibase 10.1080/00018732.2016.1198134} {\bibfield  {journal} {\bibinfo
  {journal} {Adv. Phys.}\ }\textbf {\bibinfo {volume} {65}},\ \bibinfo {pages}
  {239} (\bibinfo {year} {2016})}\BibitemShut {NoStop}%
\bibitem [{\citenamefont {Lu}\ \emph {et~al.}(2013)\citenamefont {Lu},
  \citenamefont {Bon{\v c}a},\ and\ \citenamefont {Tohyama}}]{lu2013}%
  \BibitemOpen
  \bibfield  {author} {\bibinfo {author} {\bibfnamefont {H.}~\bibnamefont
  {Lu}}, \bibinfo {author} {\bibfnamefont {J.}~\bibnamefont {Bon{\v c}a}}, \
  and\ \bibinfo {author} {\bibfnamefont {T.}~\bibnamefont {Tohyama}},\ }\href
  {http://stacks.iop.org/0295-5075/103/i=5/a=57005} {\bibfield  {journal}
  {\bibinfo  {journal} {Europhys. Lett.}\ }\textbf {\bibinfo {volume} {103}},\
  \bibinfo {pages} {57005} (\bibinfo {year} {2013})}\BibitemShut {NoStop}%
\bibitem [{\citenamefont {Mizokawa}\ \emph {et~al.}(2000)\citenamefont
  {Mizokawa}, \citenamefont {Kim}, \citenamefont {Shen}, \citenamefont {Ino},
  \citenamefont {Yoshida}, \citenamefont {Fujimori}, \citenamefont {Goto},
  \citenamefont {Eisaki}, \citenamefont {Uchida}, \citenamefont {Tagami},
  \citenamefont {Yoshida}, \citenamefont {Rykov}, \citenamefont {Siohara},
  \citenamefont {Tomimoto}, \citenamefont {Tajima}, \citenamefont {Yamada},
  \citenamefont {Horii}, \citenamefont {Yamada}, \citenamefont {Yamada},\ and\
  \citenamefont {Hirabayashi}}]{mizokawa2000}%
  \BibitemOpen
  \bibfield  {author} {\bibinfo {author} {\bibfnamefont {T.}~\bibnamefont
  {Mizokawa}}, \bibinfo {author} {\bibfnamefont {C.}~\bibnamefont {Kim}},
  \bibinfo {author} {\bibfnamefont {Z.-X.}\ \bibnamefont {Shen}}, \bibinfo
  {author} {\bibfnamefont {A.}~\bibnamefont {Ino}}, \bibinfo {author}
  {\bibfnamefont {T.}~\bibnamefont {Yoshida}}, \bibinfo {author} {\bibfnamefont
  {A.}~\bibnamefont {Fujimori}}, \bibinfo {author} {\bibfnamefont
  {M.}~\bibnamefont {Goto}}, \bibinfo {author} {\bibfnamefont {H.}~\bibnamefont
  {Eisaki}}, \bibinfo {author} {\bibfnamefont {S.}~\bibnamefont {Uchida}},
  \bibinfo {author} {\bibfnamefont {M.}~\bibnamefont {Tagami}}, \bibinfo
  {author} {\bibfnamefont {K.}~\bibnamefont {Yoshida}}, \bibinfo {author}
  {\bibfnamefont {A.~I.}\ \bibnamefont {Rykov}}, \bibinfo {author}
  {\bibfnamefont {Y.}~\bibnamefont {Siohara}}, \bibinfo {author} {\bibfnamefont
  {K.}~\bibnamefont {Tomimoto}}, \bibinfo {author} {\bibfnamefont
  {S.}~\bibnamefont {Tajima}}, \bibinfo {author} {\bibfnamefont
  {Y.}~\bibnamefont {Yamada}}, \bibinfo {author} {\bibfnamefont
  {S.}~\bibnamefont {Horii}}, \bibinfo {author} {\bibfnamefont
  {N.}~\bibnamefont {Yamada}}, \bibinfo {author} {\bibfnamefont
  {Y.}~\bibnamefont {Yamada}}, \ and\ \bibinfo {author} {\bibfnamefont
  {I.}~\bibnamefont {Hirabayashi}},\ }\href {\doibase
  10.1103/PhysRevLett.85.4779} {\bibfield  {journal} {\bibinfo  {journal}
  {Phys. Rev. Lett.}\ }\textbf {\bibinfo {volume} {85}},\ \bibinfo {pages}
  {4779} (\bibinfo {year} {2000})}\BibitemShut {NoStop}%
\end{thebibliography}

%
\end{document}